\documentstyle[twocolumn,prl,aps,epsf,psfig]{revtex}
\begin{document}                                                
\draft
\title{Gaps and Critical Temperature for Color Superconductivity}
\author{Robert D. Pisarski$^{a}$ and Dirk H. Rischke$^{b}$}
\bigskip
\address{
a) Department of Physics, Brookhaven National Laboratory,
Upton, New York 11973-5000, USA\\
b) RIKEN BNL Research Center, Brookhaven National Laboratory,
Upton, New York 11973-5000, USA
}
\date{\today}
\maketitle
\begin{abstract} 
Because of a logarithmic enhancement from soft, 
collinear magnetic gluons, in dense quark matter
the gap for a color superconducting condensate with spin zero
depends upon the QCD coupling constant $g$ not as 
$\exp(-1/g^2)$, like in BCS theory, but as $\exp(-1/g)$.
In weak coupling, the ratio of the transition temperature 
to the spin-zero gap at zero temperature 
is the same as in BCS theory. We classify the gaps
with spin one, and find that they are of the same order in $g$ as the
spin-zero gap.
\end{abstract}
\begin{narrowtext}
In cold, dense quark matter, the attractive interaction between
quarks of different colors generates color superconductivity 
\cite{bl,general,schafer0,prlett,prscalar,son,schuster,schafer,hong,hongetal}.
In this Letter we discuss in what aspects color superconductivity differs from
the classic model of Bardeen, Cooper, and Schrieffer (BCS) \cite{BCS}, and
in which aspects it resembles it.

One way in which color superconductivity differs from BCS theory
is the dependence of the condensate on the coupling constant.  
In theories with short-ranged interactions,
such as BCS theory, the gap depends upon the
coupling constant $g$ as the exponential of $1/g^2$.
We argued previously, though, that static magnetic
interactions are {\it not\/} screened to any finite order in
$g$ \cite{prlett,son}.  The scattering of quarks near the Fermi
surface is then logarithmically enhanced by the emission of
collinear, nearly static magnetic gluons, and this changes
the gap from an exponential in $1/g^2$ to one in $1/g$.

The explicit value of the gap in weak coupling
was first computed by Son \cite{son}.
Using an elegant renormalization group argument, he found that
there is an instability
at a scale $\phi_0 \sim b_0 \, \mu \, g^{-5}\, \exp(-c_0/g)$, where 
$\mu$ is the quark-chemical potential, $c_0 = 3 \pi^2/\sqrt{2}$,
and $b_0$ is a pure number. To explicitly compute
the magnitude of the spin-zero gap at zero temperature, $\phi_0$, 
it is necessary to solve a gap equation.
This was initiated by Son \cite{son}.  In this Letter we
first extend Son's analysis to estimate the
constant $b_0$. To the order in $g$ at which we work,
all of our results are manifestly gauge invariant.  
Details will be presented elsewhere \cite{wip}.

Next, we solve the gap equation at non-zero temperature, $T$,
and show that the critical temperature for the onset of
color superconductivity, $T_c$, divided by
$\phi_0$ is {\em equal\/} to the value in BCS theory \cite{BCS},
$T_c/\phi_0 \simeq 0.567 + O(g)$.

Finally, we classify the gaps for massless fermions 
with total spin $J=1$.  There are two types, longitudinal and transverse
to the direction of momentum of the quarks in the condensate, 
$\phi_1^{\parallel}$ and $\bbox{\phi}_1^{\perp}$, respectively.
In agreement with Son \cite{son}, we find that all spin-one gaps
are of the same order as the spin-zero gap:
$\phi_1/\phi_0$ is a pure number of order one.

Our results are of practical importance.
Bailin and Love assumed in their original
analysis \cite{bl} that static
magnetic interactions are screened, so that the gaps are BCS-like, and 
thus tiny, $\phi_0 \sim 10^{-3} \mu$.
Our results are only valid perturbatively, but if we 
extrapolate to strong coupling, we find that because
the constant $b_0$ is huge, the gaps can become quite large: 
as seen in fig.\ 1, $\phi_0$ peaks at $\phi_0 \sim 0.26 \, \mu$,
with a big $T_c \sim 0.15 \, \mu$. At AGS energies, 
heavy-ion collisions can probe the region of
$\mu \sim 600$ MeV and $T \sim 100$ MeV. Therefore, by triggering
on collisions in which {\it cool}, dense nuclear matter is formed,
it may be possible to observe color superconductivity.

That $J=1$ gaps are not exponentially suppressed is important for
quark stars.  At very high densities, the chemical potential of
up, down, and strange quarks are nearly equal, so the $J=0$ color-flavor
locked condensate is surely favored \cite{general}.  
At intermediate densities,
however, because of the large strange quark mass, and the requirement
of charge neutrality, these chemical potentials will differ.
This suppresses the formation of $J=0$ gaps, which are predominantly
flavor off-diagonal.  The $J=1$ gaps, however, can form between
quarks of the same flavor, and will be significant.

We follow the notations and conventions of our previous work 
\cite{prlett,prscalar}.  
For massless quarks there are four types of spin-zero condensates
\cite{bl,prlett,prscalar}: right-handed condensates
$\phi^\pm_{r,\pm}$, and left-handed condensates $\phi^\pm_{\ell,\mp}$.
The superscript refers to particles or antiparticles, while the subscript
denotes helicity.
In perturbation theory, QCD is manifestly chirally symmetric,
so that the gap equations for 
$\phi^\pm_r$ and $\phi^\pm_\ell$ are {\it identical\/}
order by order in $g^2$.
Although the magnitude of the gaps for 
$\phi_{r}$ and $\phi_{\ell}$ must
then be equal, because they are complex numbers, they differ by an
arbitrary phase.  This 
phase represents the spontaneous breaking of parity by a 
spin-zero, color superconducting gap in an instanton-free
regime \cite{prlett,prconf}.

Without loss of generality, then, we can consider only the right-handed 
gaps, denoted as $\phi^+$ and $\phi^-$, and 
take them to be real and positive.
Suppressing chiral projectors, and the color and flavor
indices \cite{color}, the gap function is 
\begin{equation}
\Phi(Q) = \phi^+(Q) \; \Lambda^+({\bf q}) 
+ \phi^-(Q) \; \Lambda^-({\bf q}) \; . \;
\label{e3}
\end{equation}
The condensate is formed from a quark, with four-momentum
$Q=(q^0,{\bf q})$, and a charge conjugate antiquark.
$\Lambda^\pm ({\bf q}) \equiv (1 \pm \gamma_0 \bbox{\gamma}
\cdot \hat{\bf q})/2$ are projectors for energy;
${\bf q} = q \, \hat{\bf q}$, $\hat{\bf q}^2 = 1$.

Including the gap, from (15) of \cite{prscalar} the quark propagator is
\begin{equation}
G(Q) = \left[
\frac{\Lambda^{+}({\bf q})}{q_0^2 - 
\epsilon^+_q\,^2}  
+\, \frac{ \Lambda^{-}({\bf q})}{q_0^2- 
\epsilon^-_q\,^2} 
\right]
 (\gamma \cdot Q - \mu \gamma_0) , 
\label{e4}
\end{equation}
where $\epsilon^\pm_q$ is
the energy of the quark relative to the Fermi surface:
\begin{equation}
\epsilon^\pm_q \equiv  \sqrt{( q \mp \mu)^2 + 
\phi^\pm(Q)^2}  \,\, .
\label{e5}
\end{equation}
The poles with $\mp \epsilon^+_q$
represent quasiparticles and their holes, 
those with $\mp \epsilon^-_q$ quasi-antiparticles and their holes
\cite{prscalar}.  At the Fermi surface, 
$q = \mu$, it takes very little
energy to excite a quasiparticle, $\epsilon^+_q = -\phi^+$,
and a lot to excite a quasi-antiparticle, $\epsilon^-_q \approx -2 \mu$.  

At one-loop order, from (A35) of \cite{prscalar}
the equation for the gap function
$\Phi(K)$ is 
\begin{equation}
\Phi(K) = \frac{2 g^2}{3}
\frac{T}{V}\sum_Q
\Delta_{\mu \nu}(K\!\!-\!\!Q)  
\gamma^\mu G_0^-(Q) \Phi(Q)  G(Q) \gamma^\nu  .
\label{e6}
\end{equation}
Here $G_0^-(Q) = 1/(\gamma \cdot Q - \mu \gamma_0)$
is the bare propagator for charge-conjugate quarks \cite{color}. 
To evaluate the Matsubara sum 
over $q^0$ we use spectral representations \cite{lebellac}. 

In the gap equation, the gluon propagator
$\Delta^{\mu \nu}$
includes the effects of ``hard dense loops'' (HDL) \cite{lebellac}.
The basic parameter of the HDL Lagrangian
is the gluon ``mass'', $m_g$; for $N_c$ colors and $N_f$
flavors of massless quarks,
\begin{equation}
m_g^2 \; = \; N_f \, \frac{g^2 \mu^2}{6 \pi^2} \; + \; 
\left( N_c + \frac{N_f}{2} \right)
\frac{g^2 T^2}{9} \; .
\label{e2}
\end{equation}
For the time being we take strict Coulomb gauge for the HDL
propagator. HDL corrections can be neglected for the quark propagator and
the quark-gluon vertex, as the quark lines are hard, $q \sim \mu$. 

We solve the gap equation by including the effects of the
superconducting state in the simplest possible
way for the quark, Eq.\ (\ref{e4}), and not at all for the gluon.
This is reasonable in weak coupling, because 
the scale of the condensate, $\phi_0 \sim \mu \, \exp(-c_0/g)$, 
is much smaller than either $\mu$ or $m_g \sim g \mu$ \cite{gap}.  

As in strong coupling BCS theory \cite{BCS}, 
$\Phi(K)$ has an imaginary part, but for small $g$ this 
can be neglected in QCD \cite{imag}. 
Consequently, the only values of the gap functions $\phi^\pm(Q)$ 
which enter into the gap equation are those on either the
quasiparticle mass shell, $\phi^+(\pm \epsilon^+_q,q)$,
or the quasi-antiparticle mass shell, $\phi^-(\pm \epsilon^-_q,q)$.
 
Gap equations for $\phi^\pm(\epsilon^\pm_k,k)$
are derived from (\ref{e6}) via projection with 
$\Lambda^\pm({\bf k})$. As is typical in models of superconductivity
\cite{BCS}, the dominant terms arise from the quasiparticle poles.
These correspond physically to scattering of
quarks near the Fermi surface. As this involves little energy transfer
between the quarks, it suffices to use the
nearly static limit of the gluon propagator.

With these approximations, denoting $\epsilon_k^+ = \epsilon_k$,
the gap equation for $\phi(k) \equiv 
\phi^+(\epsilon_k,k)$ becomes \cite{wip}
\begin{equation}
\phi(k) = \frac{g^2}{36\pi^2} \int_{\mu-\delta}^{\mu +\delta}
\frac{{\rm d}q}{\epsilon_q} \; \frac{1}{2} 
\, \ln \left( \frac{b^2 \mu^2}{\epsilon_q^2 -\epsilon_k^2} \right) 
\tanh \left(\frac{\epsilon_q}{2T} \right) \phi(q) \,\, ,
\label{e7}
\end{equation}
\begin{equation}
b = \frac{b_0}{g^5} = b_{\rm t}^2 \; b_{\rm l}^3 \; b_0' 
=  256 \; \pi^4 \left(\frac{2}{g^2 N_f}\right)^{5/2} b_0'\;,
\label{e8}
\end{equation}
where $ b_{\rm t} = 4 \sqrt{2}\, \mu/(\sqrt{3 \pi} \,m_g)$,
and $b_{\rm l}  = 2 \, \mu/(\sqrt{3} \, m_g)$.
The logarithm $\sim \ln[1/(\epsilon_q^2 - \epsilon_k^2)]$ arises
from the cut term in the spectral density of a nearly static transverse
gluon \cite{prlett,son}.
In the gap equation, there are also terms $\sim \ln(1/g)$ which
arise from the non-static transverse gluons and from static
longitudinal gluons; these produce the
constants $b_{\rm t}$ and $b_{\rm l}$, respectively.
In addition, there are terms $\sim 1$ in the gap equation which
contribute to the constant $b_0'$; we did not compute these terms.
In deriving (\ref{e7}) we assume that $\epsilon_k, \epsilon_q < \mu$, 
so we introduce a cut-off $\delta$ on the $q$-integration;
the final result is independent of $\delta$.

At $T=0$, an approximate solution of (\ref{e7}) is \cite{wip}
\begin{eqnarray} \label{e10}
\phi(k) & = & \phi_0 \, \sin (\bar{g}\, y_k) \,\, , \\
\bar{g} \equiv  \frac{g}{3\sqrt{2}\pi}  & , & \;\;\;
y_k  \equiv \ln \left( \frac{2b\mu}{|k-\mu|+\epsilon_k} \right)\;,\nonumber
\end{eqnarray}
where $\phi_0$ denotes the value of the condensate at 
the Fermi surface, $k = \mu$. (This is similar, but not identical
to the solution of \cite{son,schafer,hongetal}.)
As $y_\mu = \ln(2b\mu/\phi_0)$, Eq.\ (\ref{e10}) requires 
$\bar{g}\, y_\mu = \pi/2$, {\it i.e.},
\begin{equation} \label{e1}
\phi_0 = 2\, b \mu \, \exp\left(-\frac{\pi}{2 \bar{g}}\right)\;.
\end{equation}
Our results for $c_0$ and the prefactor $1/g^5$ are in agreement
with Son \cite{son}. The constants 
$b_{\rm t}$ and $b_{\rm l}$ are the same found in an independent
analysis by Sch\"afer and Wilczek \cite{schafer}; see also
Hong {\it et al.} \cite{hongetal}.

In BCS-like theories with zero-ranged interactions, such as
Nambu--Jona-Lasinio (NJL) models \cite{general}, all 
particle pairs around the Fermi surface contribute with
{\it equal\/} weight to build up the BCS-logarithm, so that the gap 
function is constant: $\sim g^2 \int
{\rm d}q / \epsilon_q \simeq g^2 \ln(2\delta/\phi_0)$,
with solution $\phi_0 \sim 2 \delta \, \exp(-1/g^2)$.
In a model where fermions interact with
scalar bosons of mass $M_s \sim g \mu$ \cite{prscalar},
scattering of particle pairs through small angles is {\it favored}.
The collinear singularity is cut off by $M_s \neq 0$, so that 
logarithmic factors of $\sim \ln (\mu/M_s) \sim \ln(1/g)$ appear in the
gap equation.

In QCD, the scattering of quark pairs through small angles
is again favored. If the exchanged gluon is electric, the 
collinear singularity is cut off by the Debye mass, $\sqrt{3}\,
m_g$. This produces $\ln (1/g)$ terms which 
contribute to the prefactor $1/g^5$ in $b$, Eq.\ (\ref{e8}). 
If the exchanged gluon
is magnetic, the collinear singularity is only cut off by the
difference in energies between the incoming and outgoing pairs.
In the gap equation (\ref{e7}), this
generates the logarithmic enhancement factor $\sim \ln[1/(\epsilon_q^2 -
\epsilon_k^2)]$. The dependence of the
gap function on $\epsilon_k$ is then not negligible. 
Quasiparticles with momenta exponentially close to the Fermi surface,
$\epsilon_q \sim b \mu \, \exp(-c/g)$, dominate the integral,
with a contribution which is enhanced by $ \ln(b\mu/\epsilon_q) \sim c/g$.
The gap function $\phi(q)$ is weighted towards these pairs, as
$\phi(q)/\phi_0 \sim \sin(\pi c/2c_0) \sim 1$.
For quasiparticles which are not exponentially close to the Fermi
surface, $\epsilon_q \sim \mu$ and $c\sim g$, the gap function is
down by $\phi(q)/\phi_0 \sim g$ \cite{imag}.

The temperature dependence of the condensate can be computed from
Eq.\ (\ref{e7}) as follows. We assume that the temperature $T$ is 
of the order of the gap at zero temperature, $\phi_0$.
Let us introduce a dimensionless parameter $\kappa \gg 1$, and divide
the integration region into $ \epsilon_q\geq \kappa \phi_0$ and
$\epsilon_q < \kappa \phi_0$. 
Away from the Fermi surface, $\epsilon_q \gg \phi_0$, 
the Fermi--Dirac distribution becomes a Boltzmann distribution, so
$\tanh (\epsilon_q/2T) \simeq 1$, and thermal effects are negligible.
Near the Fermi surface, the thermal factor $\tanh (\epsilon_q/2T)$ 
cuts off any singularity, even at the critical temperature, $T_c$,
when $\phi(q) \rightarrow 0$.
Then the gap function is the same as (\ref{e10}) for 
$\epsilon_k \gg \kappa \phi_0$, and a constant for
$\epsilon_k \ll \kappa \phi_0$.  Matching the two regions at
$\kappa \phi_0$, and then sending $\kappa \rightarrow \infty$,
we derive the condition 
\begin{equation} \label{e12}
\int_0^\infty {\rm d}|q-\mu| \left[ \frac{1}{\epsilon_q} \,
\tanh\left(\frac{\epsilon_q}{2T}\right) - \frac{1}{\epsilon_q^0}
\right] = 0\;,
\end{equation}
where $\epsilon_q = \sqrt{(q-\mu)^2 +\phi^2(T)}$, with $\phi(T)$ 
the gap at the Fermi surface at a temperature $T$, and
$\epsilon_q^0 = \sqrt{(q-\mu)^2 +\phi_0^2}$. 
This is correct to leading order in $g$.
Equation (\ref{e12}) implicitly determines the function $\phi(T)/\phi_0$; it is
{\em identical\/} to that obtained in BCS theory in
weak coupling \cite{BCS,wip}. In particular, the ratio of the critical
temperature to the zero-temperature gap is the same as in BCS, 
$T_c/\phi_0 = \zeta/2 +O(g)$, where the constant 
$\zeta=2\, e^\gamma /\pi\simeq 1.13$. 
($\gamma \simeq 0.577$ is the Euler-Mascheroni constant.)

Following the classification of \cite{prlett,prscalar},
for massless quarks a spin-one condensate has the form
\begin{equation}
\sum_{h, e} \left( \bbox{\phi}^{\parallel e}_h(Q)
\cdot \hat{\bf q} + \bbox{\phi}^{\perp e}_h(Q)
\cdot {\bf P}({\bf q}) \cdot \bbox{\gamma}
\right) {\cal P}_h \, \Lambda^{e}({\bf q})  \;,
\end{equation}
where the sum runs over chiralities, $h= r,\ell$, and energies,
$e=\pm$. ${\cal P}_{r,\ell} = (1 \pm \gamma_5)/2$ is the 
chiral projector, and ${\bf P}({\bf q})={\bf 1} - \hat{\bf q} \hat{\bf q}$
a projector onto the subspace orthogonal to ${\bf q}$.
Because a spin-one condensate is a three-vector,
there are 12 types of condensates, four longitudinal,
$\phi_{1,h}^{\parallel e} \equiv \bbox{\phi}^{\parallel e}_{h} \cdot 
\hat{\bf q}$, and eight transverse, $\bbox{\phi}_{1,h}^{\perp e} \equiv
\bbox{\phi}^{\perp e}_h \cdot {\bf P} ({\bf q})$. 
This classification is equivalent to that of \cite{bl}. While 
the spin-zero gaps are symmetric in the simultaneous
interchange of color and flavor indices \cite{bl,prlett,prscalar}, the
longitudinal gaps $\phi_{1,h}^{\parallel e}$ are
antisymmetric. The transverse gaps fulfill a more complicated
relationship,
$(\bbox{\phi}^{\perp \pm}_{r, \ell})^T = - \bbox{\phi}^{\perp \mp}_{\ell,r}$.
The spin-zero gaps and the longitudinal spin-one gaps do not mix 
quarks of different chirality; the transverse spin-one gaps do,
and thus break chiral symmetry.
The gap equations can be constructed as in the spin-zero case
\cite{wip}. We find that both the longitudinal 
as well as the transverse gaps fulfill the same gap equation as 
the spin-zero gaps, with identically
the same solution as in (\ref{e1}), except that
the constant analogous to $b_0'$ (which we do not compute) may differ. 

In the static limit,
gauge dependent terms in the gluon propagator $\Delta^{\mu \nu}(P)$ are 
$\sim p^\mu p^\nu/p^2$  \cite{lebellac}.
These terms contribute to the gap equation, but neither to $c_0$,
the power of $g$ in the prefactor, $b_{\rm t}$, nor $b_{\rm l}$.
They do appear to contribute to the undetermined constant $b_0'$, 
but we suggest that in the end, $b_0'$ is gauge invariant.
There are other effects which contribute to $b_0'$ \cite{wip}.
One-loop diagrams with a {\em soft}, transverse HDL gluon propagator
renormalize the quark \cite{wave} and gluon wave functions, and the 
quark-gluon vertex. 
Other contributions arise from the influence of the condensate 
on the gluon propagator \cite{gap}, and the admixture of 
quasi-antiparticle modes in the quasiparticle gap equation. 
It is important to calculate $b_0'$, since its
numerical value determines 
exactly {\it which\/} patterns of symmetry breaking are favored.

\begin{figure} 
\vspace*{1cm}
\psfig{figure=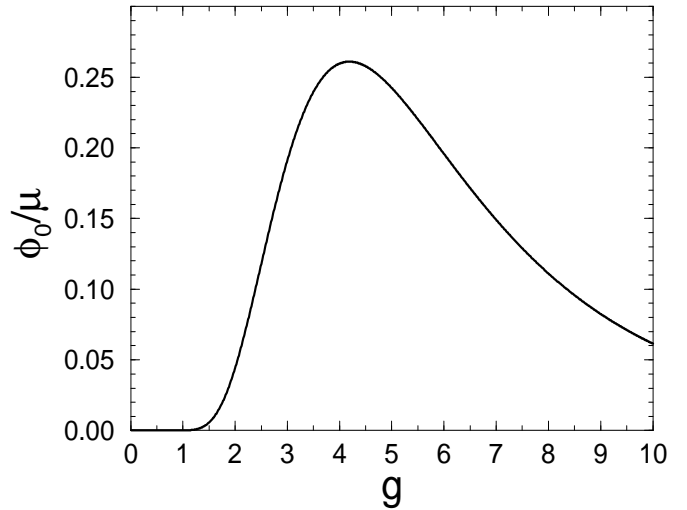,width=2.in,height=2.8in,angle=-90}
\vspace*{-1.5cm}
\caption{$\phi_0/\mu$ as function of $g$, for $b_0' = 1$.}
\end{figure}

While the results which we have derived are rigorously valid only
in weak coupling, it is interesting to plot $\phi_0/\mu$ as a function of
$g$, fig.\ 1. We take $N_f=2$ 
(note that $b_0 \sim 1/N_f^{5/2}$).  From 
(\ref{e8}), $\phi_0/\mu$ is proportional to $b_0'$;
in fig.\ 1 we set this undetermined constant equal to 1. Equation (\ref{e1})
has the form of a semiclassical tunneling probability,
including a prefactor from five zero modes.  Because of the 
``zero modes'', the gap function peaks at a value of $\phi_0/\mu \sim 0.26$
when $g \sim 4.2$.  

Extending the 
picture of Sch\"afer and Wilczek \cite{schafer0},
we view quark matter as a color superconducting ``liquid'',
and hadronic matter as 
a color superconducting ``vapor''. From \cite{prlett} there is 
a first-order phase transition between these liquid and vapor phases
at $\mu=\mu_c$ and $T=0$. Then perhaps at several times
nuclear matter density, the liquid phase occurs
at the maximum of $\phi_0/\mu$, and the vapor phase at larger $g$,
providing a qualitative explanation for the smallness of the 
analogous gaps in hadronic matter \cite{general}.

We conclude by using (\ref{e6}) to estimate the validity of
perturbation theory.  Perturbative calculations break down
when $m_g \simeq \mu$ or $T$. For $N_c= 3$ and $N_f = 2$,
at $T \neq 0$ and $\mu =0$, 
$m_g = T$ when the QCD fine structure constant is tiny,
$\alpha_s \equiv g^2/4 \pi \sim 0.18$.  
In contrast, at $\mu \neq 0$ and 
$T = 0$, $m_g = \mu$ when $\alpha_s$ is much larger,
$\alpha_s \sim 2.4$.  This suggests to us that while
perturbation theory is not a good approximation for hot 
quark-gluon matter \cite{temp}, it may well be a reasonable
guide to understanding dense quark matter, as long as it is cold,
$T < 0.3 \mu$.

This work was supported in part by DOE grant DE-AC02-98CH10886.
We thank M.\ Alford, J.\ Berges, W.\ Brown,
V.\ Emery, D.K.\ Hong, S.D.H.\ Hsu, J.T.\ Liu,
V.N.\ Muthukumar, K.\ Rajagopal, H.C.\ Ren, T.\ Sch\"afer,
D.\ Son, and F.\ Wilczek for enlightening discussions.
We especially thank T.\ Sch\"afer for discussions 
on the ratio $T_c/\phi_0$.
D.H.R.\ thanks RIKEN, 
BNL and the U.S.\ Department
of Energy for providing the facilities essential for
the completion of this work.
\end{narrowtext}

\end{document}